# Axisymmetric electrovacuum spacetimes with a translational Killing vector at null infinity

J. Bičák[*]

*Department of Theoretical Physics, Faculty of Mathematics and Physics, Charles University, V Holešovičkách 2, 180 00 Prague 8, Czech Republic*

A. Pravdová[†]

*Mathematical Institute, Academy of Sciences, Žitná 25, 115 67 Prague 1, Czech Republic*

(August 31, 2018)

By using the Bondi-Sachs-van der Burg formalism we analyze the asymptotic properties at null infinity of axisymmetric electrovacuum spacetimes with a translational Killing vector and, in general, an infinite "cosmic string" (represented by a conical singularity) along the axis. Such spacetimes admit only a local null infinity. There is a non-vanishing news function due to the existence of the string even though there is no radiation.

We prove that if null infinity has a smooth compact cross section and the Bondi mass is non-vanishing, then the translational Killing vector must be timelike and the spacetime is stationary. The other case in which an additional symmetry of axisymmetric spacetimes admits compact cross sections of null infinity is the boost symmetry, which leads to radiative spacetimes representing "uniformly accelerated objects". These cases were analyzed in detail in our previous works. If the translational Killing vector is spacelike or null, corresponding to cylindrical or plane waves, some complete generators of null infinity are "singular" but null infinity itself can be smooth apart from these generators.

As two explicit examples of local null infinity, Schwarzschild spacetime with a string and a class of cylindrical waves with a string are discussed in detail in the Appendix.

PACS numbers: 04.20.Ha, 04.20.Jb, 04.30.-w, 04.40.Nr

## I. INTRODUCTION AND SUMMARY

Recently we studied symmetries compatible with asymptotic flatness and admitting gravitational and electromagnetic radiation by using the Bondi-Sachs-van der Burg formalism [1]. We have shown that in axially symmetric electrovacuum spacetimes in which at least locally a smooth null infinity in the sense of Penrose exists, the only second allowable symmetry is either the translational symmetry or the boost symmetry. The boost-rotation symmetric spacetimes are radiative; they describe "uniformly accelerated charged, spinning particles or black holes". In [1] the emphasis was on these spacetimes. For example, the general functional forms of gravitational and electromagnetic news functions and of the mass at null infinity of the boost-rotation symmetric spacetimes have been obtained.

Here we concentrate on the other case: axisymmetric electrovacuum spacetimes which admit asymptotically translational Killing fields at null infinity. Our analysis can thus be considered as an extension of the recent extensive study of asymptotically translational Killing vectors at spatial infinity by Beig and Chruściel [2]. (In subsequent work [3] these authors gave a complete classification of all connected isometry groups at spatial infinity of asymptotically flat, asymptotically vacuum spacetimes with timelike ADM four-momentum.) We notice, however, that Beig and Chruściel [2] did not use the field equations in their analysis whereas we assume the Einstein-Maxwell equations to be satisfied.

In general we suppose that null infinity exists only locally, i.e., it need not admit smooth cross sections. In particular, we assume that in addition to a bounded system there may exist an infinite thin cosmic string (represented by a conical singularity) along the symmetry axis $\theta = 0, \pi$ so that two generators of null infinity are missing. As it is shown in [4],

---

[*]Email address: bicak@mbox.troja.mff.cuni.cz

[†]Email address: pravdova@mbox.troja.mff.cuni.cz and pravdova@math.cas.cz

[5] and analyzed in [1], there is a non-vanishing news function due to the presence of the string even though spacetime may be non-radiative.

Another case in which we expect some generators of null infinity to be missing is that of cylindrical waves. These admit a spacelike translational Killing vector and, in an appropriate coordinate system, we expect that null infinity will be singular at $\theta = \pi/2$. In our previous work [1] we did not consider this case. We notice, however, that the behaviour of cylindrical waves at null infinity was analyzed in detail both within four-dimensional spacetime and the three-dimensional reduction thereof in [6], [7]. Here we consider the translational Killing field to be of any character, i.e., it may be timelike, spacelike, or null, i.e., we admit cylindrical waves.

In the next section we recapitulate briefly some basic equations and relations from [1] which will be needed (an acquaintance with the detailed work of [1] would still be very helpful in understanding the present paper). In particular, we summarize the results obtained in [1] for the case in which the second allowable symmetry in axisymmetric electrovacuum spacetimes is a translation. In Section III the translationally invariant spacetimes with, in general, an infinite cosmic string along the axis of symmetry are studied in detail. Using the Bondi-Sachs-van der Burg formalism (with the field equations as corrected in Appendix A in [1]), we expand both the Killing equations and the corresponding Lie equations for electromagnetic field in further orders of $r^{-k}$. Asymptotic expansions of the Killing vector and functions determining the metric and electromagnetic field are given in general under the presence of the string along the $z$-axis and, in simplified explicit forms, in cases without the string. The complete metric and electromagnetic field tensors are displayed in Appendix A.

In Section IV the norm of the translational Killing field is calculated and the theorem is proved that, "if the Bondi mass is non-vanishing and null infinity admits smooth compact cross sections, then the translational Killing vector has to be timelike, i.e., the spacetime has to be stationary" (see Theorem 2). The resulting expansions should be useful in those problems in which the asymptotic forms of stationary metrics in Bondi-type coordinates are used. The spacelike and null translational Killing vector correspond to cylindrical or plane waves respectively. Complete generators of null infinity are then singular at a given $\theta_0$ (see Theorem 1) although local smooth null infinity may exist for other values of $\theta$. The only axisymmetric radiative spacetimes with an additional symmetry which admit compact smooth cross sections of null infinity are the boost-rotation symmetric spacetimes investigated in [1]. Their general structure was geometrically analyzed in [8].

In Appendix B two examples of translationally invariant axisymmetric vacuum spacetimes with a local null infinity are studied explicitly: the Schwarzschild spacetime with an infinite string and cylindrical waves with an infinite string along the axis of symmetry. In the first case, null infinity is singular for $\theta = 0, \pi$, in the second case it is singular for both $\theta = 0, \pi$ and $\theta = \pi/2$. Due to the presence of string the news function is non-vanishing in both cases.

## II. AXISYMMETRIC ELECTROVACUUM SPACETIMES WITH ANOTHER SYMMETRY

In [1] we considered electrovacuum spacetimes with axial Killing vector $\partial/\partial\phi$ which admit at least the "piece of $\mathcal{I}^+$" in the sense that one can introduce the Bondi-Sachs coordinates $\{u, r, \theta, \phi\} \equiv \{x^0, x^1, x^2, x^3\}$ in which the metric satisfying the Einstein-Maxwell equations reads

$$
\begin{aligned}
ds^2 = & \left(\frac{V}{r}e^{2\beta} - r^2 e^{2\gamma}U^2 \cosh 2\delta - r^2 e^{-2\gamma}W^2 \cosh 2\delta - 2r^2 UW \sinh 2\delta\right) du^2 \\
& + 2e^{2\beta} du dr + 2r^2(e^{2\gamma}U \cosh 2\delta + W \sinh 2\delta) du d\theta + 2r^2(e^{-2\gamma}W \cosh 2\delta + U \sinh 2\delta) \sin\theta \, du d\phi \\
& - r^2 \left[\cosh 2\delta(e^{2\gamma}d\theta^2 + e^{-2\gamma}\sin^2\theta \, d\phi^2) + 2\sinh 2\delta \sin\theta \, d\theta d\phi\right] \, .
\end{aligned} \tag{1}
$$

Here the metric functions at large $r$ with $u$, $\theta$, $\phi$ fixed have expansions (in which the field equations are used) of the form (cf. Eqs. (A20)–(A31) in [1])

$$
\begin{aligned}
\gamma &= \frac{c}{r} + (C - \tfrac{1}{6}c^3 - \tfrac{3}{2}cd^2)\frac{1}{r^3} + O(r^{-4}) \, , \\
\delta &= \frac{d}{r} + (H - \tfrac{1}{6}d^3 + \tfrac{1}{2}c^2 d)\frac{1}{r^3} + O(r^{-4}) \, , \\
\beta &= -\tfrac{1}{4}(c^2 + d^2)\frac{1}{r^2} + O(r^{-4}) \, , \\
U &= -(c_{,\theta} + 2c \cot\theta)\frac{1}{r^2} + [2N + 3(cc_{,\theta} + dd_{,\theta}) + 4(c^2 + d^2)\cot\theta]\frac{1}{r^3} \\
&\quad + \tfrac{1}{2}[3(C_{,\theta} + 2C \cot\theta) - 6(cN + dP) - 4(2c^2 c_{,\theta} + cdd_{,\theta} + c_{,\theta} d^2)
\end{aligned}
$$

$$-8c(c^2+d^2)\cot\theta+2(\epsilon e-f\mu)]\frac{1}{r^4}+O(r^{-5})\ , \tag{2}$$

$$W=-(d_{,\theta}+2d\cot\theta)\frac{1}{r^2}+[2P+2(c_{,\theta}d-cd_{,\theta})]\frac{1}{r^3}$$
$$+\tfrac{1}{2}[3(H_{,\theta}+2H\cot\theta)+(cP-dN)-4(2d^2d_{,\theta}+cdc_{,\theta}+c^2d_{,\theta})$$
$$-8d(c^2+d^2)\cot\theta+2(\mu e+\epsilon f)]\frac{1}{r^4}+O(r^{-5})\ ,$$

$$V=r-2M-[N_{,\theta}+N\cot\theta-\tfrac{1}{2}(c^2+d^2)$$
$$-(c_{,\theta}+2c\cot\theta)^2-(d_{,\theta}+2d\cot\theta)^2-(\epsilon^2+\mu^2)]\frac{1}{r}$$
$$-\tfrac{1}{2}[C_{,\theta\theta}+3C_{,\theta}\cot\theta-2C+6N(c_{,\theta}+2c\cot\theta)+6P(d_{,\theta}+2d\cot\theta)$$
$$+4(2cc_{,\theta}^2+3c_{,\theta}dd_{,\theta}-cd_{,\theta}^2)+8(2c_{,\theta}d^2+3c^2c_{,\theta}+cdd_{,\theta})\cot\theta$$
$$+16c(c^2+d^2)\cot^2\theta+2\epsilon(e_{,\theta}+e\cot\theta)-2\mu(f_{,\theta}+f\cot\theta)]\frac{1}{r^2}+O(r^{-3})\ ,$$

and the electromagnetic field reads

$$F_{01}=-\frac{\epsilon}{r^2}+(e_{,\theta}+e\cot\theta)\frac{1}{r^3}+O(r^{-4})\ ,$$
$$F_{02}=X+(\epsilon_{,\theta}-e_{,u})\frac{1}{r}-\{[E+\tfrac{1}{2}(ec+fd)]_{,u}+\tfrac{1}{2}(e_{,\theta}+e\cot\theta)_{,\theta}\}\frac{1}{r^2}+O(r^{-3})\ ,$$
$$F_{03}=\left\{Y-\frac{f_{,u}}{r}-\{[F+\tfrac{1}{2}(ed-fc)]_{,u}\}\frac{1}{r^2}+O(r^{-3})\right\}\sin\theta\ , \tag{3}$$
$$F_{12}=\frac{e}{r^2}+(2E+ec+fd)\frac{1}{r^3}+O(r^{-4})\ ,$$
$$F_{13}=\left[\frac{f}{r^2}+(2F+ed-fc)\frac{1}{r^3}+O(r^{-4})\right]\sin\theta\ ,$$
$$F_{23}=\left[-\mu-(f_{,\theta}+f\cot\theta)\frac{1}{r}+O(r^{-2})\right]\sin\theta\ .$$

The "coefficients" $c$, $d$, $C$, $H$, $N$, $P$, $M$, $e$, $f$, $E$, $F$, $X$, $Y$, $\epsilon$, $\mu$ are functions of $u$ and $\theta$. The mass aspect $M(u,\theta)$ is connected with the gravitational news functions $c_{,u}$ and $d_{,u}$ and with the electromagnetic news functions $X$ and $Y$ by the relation

$$M_{,u}=-(c_{,u}^2+d_{,u}^2)-(X^2+Y^2)+\tfrac{1}{2}(c_{,\theta\theta}+3c_{,\theta}\cot\theta-2c)_{,u}\ . \tag{4}$$

The field equations also imply relations which will be further needed (cf. Eqs. (A32)–(A37), (A46)–(A47) in [1])

$$3N_{,u}=-M_{,\theta}-2c(\partial_\theta+2\cot\theta)c_{,u}-2d(\partial_\theta+2\cot\theta)d_{,u}-(cc_{,u})_{,\theta}-(dd_{,u})_{,\theta}-2(\epsilon X+\mu Y)\ , \tag{5}$$
$$3P_{,u}=\tfrac{1}{2}\partial_\theta(\partial_\theta+\cot\theta)(d_{,\theta}+2d\cot\theta)+2c(\partial_\theta+2\cot\theta)d_{,u}-2d(\partial_\theta+2\cot\theta)c_{,u}$$
$$+(cd_{,u})_{,\theta}-(dc_{,u})_{,\theta}-2(\epsilon Y-\mu X)\ , \tag{6}$$
$$e_{,u}=\tfrac{1}{2}\epsilon_{,\theta}-(cX+dY)\ , \tag{7}$$
$$f_{,u}=-\tfrac{1}{2}\mu_{,\theta}-(-cY+dX)\ , \tag{8}$$
$$2e_{,u}=\epsilon_{,\theta}-2(cX+dY)\ , \tag{9}$$
$$2f_{,u}=-\mu_{,\theta}-2(-cY+dX)\ , \tag{10}$$
$$4E_{,u}=-(\partial_\theta+2\cot\theta)[(\partial_\theta-\cot\theta)e+2(c\epsilon+d\mu)]\ , \tag{11}$$
$$4F_{,u}=-(\partial_\theta+2\cot\theta)[(\partial_\theta-\cot\theta)f+2(d\epsilon-c\mu)]\ , \tag{12}$$
$$4C_{,u}=2(c^2-d^2)c_{,u}+4dcd_{,u}+2cM+d(\partial_\theta+\cot\theta)(d_{,\theta}+2d\cot\theta)-(\partial_\theta-\cot\theta)N+2(eX-fY)\ , \tag{13}$$
$$4H_{,u}=-2(c^2-d^2)d_{,u}+4dcc_{,u}+2dM-c(\partial_\theta+\cot\theta)(d_{,\theta}+2d\cot\theta)-(\partial_\theta-\cot\theta)P+2(eY+fX)\ . \tag{14}$$

Since we admit also spacetimes with only "local" $\mathcal{I}^+$, we assume the above equations to be in general true in some open interval of $\theta$. In particular, the "axis of symmetry" ($\theta=0$, $\pi$) may contain nodal singularities ("cosmic strings"), or $\mathcal{I}^+$ may be singular at some $\theta=\theta_0\neq 0$, $\pi$ which will be the case with cylindrical waves, as we shall see in the following. Then, of course, some of generators of $\mathcal{I}^+$ are missing (being "singular").

Now we assumed in [1] that another Killing field $\eta^\alpha$ exists which forms with $\partial/\partial\phi$ a two-parameter group. We decompose this field into the standard null tetrad field $\{k^\alpha, m^\alpha, t^\alpha, \bar{t}^\alpha\}$ where

$$k_\alpha = \left[1\ ,\ 0\ ,\ 0\ ,\ 0\right], \quad m_\alpha = \left[\tfrac{1}{2}Vr^{-1}e^{2\beta}\ ,\ e^{2\beta}\ ,\ 0\ ,\ 0\right],$$

$$t_\alpha = \tfrac{1}{2}r(\cosh 2\delta)^{-\tfrac{1}{2}}\Big[(1+\sinh 2\delta)e^\gamma U + \cosh 2\delta e^{-\gamma}W + i[(1-\sinh 2\delta)e^\gamma U - \cosh 2\delta e^{-\gamma}W], \tag{15}$$

$$0\ ,\ -(1+\sinh 2\delta + i(1-\sinh 2\delta))e^\gamma\ ,\ -(1-i)\cosh 2\delta \sin\theta e^{-\gamma}\Big].$$

We thus write

$$\eta^\alpha = Ak^\alpha + Bm^\alpha + \tilde{f}(t_R^\alpha + t_I^\alpha) + \tilde{g}(t_R^\alpha - t_I^\alpha), \tag{16}$$

where $A$, $B$, $\tilde{f}$, $\tilde{g}$ are general functions of $u$, $r$, $\theta$, the subscripts $R$ and $I$ denote the real and imaginary parts. The Killing equations,

$$\mathcal{L}_\eta g_{\alpha\beta} = 0, \tag{17}$$

first imply (for $\alpha = \beta = 1$) that function

$$B = B(u, \theta) \tag{18}$$

is independent of $r$. We solve Killing equations asymptotically assuming that functions $A$, $\tilde{f}$, $\tilde{g}$ can be expanded in powers of $r^{-k}$. Eqs. (17) imply that the leading terms are proportional to $r$ (cf. (18) in [1]), so that we can write

$$A = A^{(-1)}r + A^{(0)} + \frac{A^{(1)}}{r} + \frac{A^{(2)}}{r^2} + O(r^{-3}),$$

$$\tilde{f} = f^{(-1)}r + f^{(0)} + \frac{f^{(1)}}{r} + \frac{f^{(2)}}{r^2} + O(r^{-3}), \tag{19}$$

$$\tilde{g} = g^{(-1)}r + g^{(0)} + \frac{g^{(1)}}{r} + \frac{g^{(2)}}{r^2} + O(r^{-3}),$$

where $A^{(k)}$, $f^{(k)}$, $g^{(k)}$ are functions of $u$ and $\theta$.

By solving Killing equations (17) in the leading orders it is proved in [1] (see the Theorem above Eq. (30) therein) that the general asymptotic form of the Killing vector $\eta^\alpha$ is

$$\eta^\alpha = [-ku\cos\theta + \alpha(\theta)\ ,\ kr\cos\theta + O(r^0)\ ,\ -k\sin\theta + O(r^{-1})\ ,\ O(r^{-1})], \tag{20}$$

where $k$ is a constant, $\alpha$ – an arbitrary function of $\theta$. When $k \neq 0$, one can put $\alpha = 0$ and $k = 1$. The field $\eta^\alpha$ is then asymptotically the boost Killing vector which generates the Lorentz transformations along the axis of axial symmetry. The case of the boost Killing vector is analyzed in detail in Section IV in [1].

When $k = 0$, the field (20) is asymptotically the supertranslational Killing field. However, as shown in Section III in [1] (by considering the Killing equations (17) in the higher orders in $r^{-k}$, the Lie equations for $F_{\mu\nu}$ and some of the Einstein-Maxwell equations) the field turns out to be, in fact, the translational Killing field. In general a straight infinite thin cosmic string along the $z$-axis is permitted in [1] which is characterized by constant $\mathcal{C}$ (denoted by C in [1]) where

$$\mathcal{C} \in (0,\ 1\rangle, \tag{21}$$

so that in the weak-field limit $\mathcal{C} = 1 - 4\mu$, $\mu$ being the mass per unit length of the string; the deficit angle due to the string is given by $2\pi(1-\mathcal{C})$ – see [1], in particular Appendix D, for more details. The following results have been obtained in the translational case in [1]. Functions $B(u, \theta)$, $A^{(k)}$, $f^{(k)}$, $g^{(k)}$, determining the asymptotic form of the translational Killing field and the leading metric and field functions, are given by

$$A^{(-1)} = f^{(-1)} = g^{(-1)} = g^{(0)} = f^{(1)} = 0,$$

$$B = B(\theta) = a\sin\theta\left(\frac{\sin\theta}{\cos\theta+1}\right)^\mathcal{C} + b\sin\theta\left(\frac{\sin\theta}{\cos\theta+1}\right)^{-\mathcal{C}}, \quad a,\ b = \text{const.},$$

$$A^{(0)} = \tfrac{1}{2}(B_{,\theta\theta} + B_{,\theta}\cot\theta + B), \quad f^{(0)} = -B_{,\theta}, \quad A^{(1)} = A^{(1)}(\theta), \quad g^{(1)} = B(d_{,\theta} + 2d\cot\theta) - B_{,\theta}d, \tag{22}$$

$$c = \frac{u}{2B}(B_{,\theta\theta} - B_{,\theta}\cot\theta) = u\frac{\mathcal{C}^2-1}{2\sin^2\theta}, \quad d = d(\theta),$$

$$M = -uc_{,u}^2 - A^{(1)}B^{-1},$$

$$X = Y = 0, \quad \epsilon = \epsilon_0 B^{-2}, \quad \mu = \mu_0 B^{-2}, \quad e = -\epsilon_0 B_{,\theta}B^{-3}u + e_1(\theta), \quad f = \mu_0 B_{,\theta}B^{-3}u + f_1(\theta).$$

(The expressions for $M$, $\epsilon$ and $\mu$ do not appear in [1] explicitly but they follow immediately from Eqs. (56), (67), (72), (74), (70), (73) in [1].) The news function, $c_{,u}$, is non-vanishing if the cosmic string occurs along the axis, i.e., if $\mathcal{C} \neq 1$, and it is singular for $\theta = 0$, $\pi$; the Weyl tensor is, of course, non-radiative (cf. [4]).

To see the meaning of constants $a$, $b$, consider translations along the $z$-axis and $t$-axis in cylindrical coordinates in spacetimes with a straight cosmic string along the $z$-axis (see Appendix D in [1]):

$$\zeta^\mu_{(z)} = [0, 0, a_0, 0] \,, \quad \zeta^\mu_{(t)} = [b_0, 0, 0, 0] \,, \tag{23}$$

$a_0$, $b_0$ = const. Going over to Bondi's coordinates in which in first orders the Killing field is (see Eq. (58) in [1])

$$\eta^\mu = \left[ B(\theta) \,, \tfrac{1}{2}(B_{,\theta\theta} + B_{,\theta} \cot\theta) + O(r^{-1}) \,, -B_{,\theta}\frac{1}{r} + B_{,\theta}\frac{c}{r^2} + O(r^{-3}) \,, B_{,\theta}\frac{d}{r^2}\sin\theta + O(r^{-3}) \right] \,, \tag{24}$$

we find that constants $a$, $b$ are related to $a_0$, $b_0$ by

$$a = \frac{(b_0 + a_0)\chi^\mathcal{C}}{2\mathcal{C}} \,, \quad b = \frac{(b_0 - a_0)\chi^{-\mathcal{C}}}{2\mathcal{C}} \,; \tag{25}$$

$b_0 = 0$ gives a translation along the $z$-axis, $a_0 = 0$ - along the $t$-axis. Different values of constant parameter $\chi$ just correspond to Bondi's coordinates which are boosted along the $z$-axis with velocity $v = -\tanh(\ln \chi^{-\mathcal{C}})$. In the following we choose the unboosted system by putting $\chi = 1$. (In Appendix D in [1] we left $\chi$ arbitrary.) Introducing function $q$ by

$$q = \frac{\sin\theta}{2\mathcal{C}} \left[ \left(\frac{\sin\theta}{\cos\theta + 1}\right)^\mathcal{C} + \left(\frac{\sin\theta}{\cos\theta + 1}\right)^{-\mathcal{C}} \right] \,, \tag{26}$$

we can simplify function $B$ entering the Killing field:

$$B = B_T + B_Z \,, \tag{27}$$

where

$$B_T = b_0 q \,, \quad B_Z = \frac{a_0}{\mathcal{C}}(q_{,\theta}\sin\theta - q\cos\theta) \,, \tag{28}$$

"T" corresponding to translations along the $t$-axis, "Z" – along the $z$-axis. The above expressions still simplify significantly if there is no string along the axis, i.e., if $\mathcal{C} = 1$:

$$B = -a_0 \cos\theta + b_0 \,, \quad q = 1 \,. \tag{29}$$

### III. ASYMPTOTICALLY TRANSLATIONAL KILLING VECTORS

Starting from the metric (1)–(2), the electromagnetic field (3), and the Killing field in the form (15), (16), (19) in which the first coefficients are determined by (22), we shall now expand the Killing equations (17) in higher orders in $r^{-1}$. We find the following restrictions on the expansion coefficients $A^{(k)}$, $f^{(k)}$, $g^{(k)}$ and functions entering the metric and the field (in which, for the sake of brevity of the equations, we do not substitute expressions for non-vanishing quantities $B$, $A^{(0)}$, $f^{(0)}$, $g^{(1)}$, $M$, $\epsilon$, $\mu$ from (22) but whenever the term $c_{,\theta} + 2c\cot\theta$ appears we use the fact that as a consequence of (22) it vanishes):

$$\mathcal{L}_\eta g_{00} = 0 \quad (r^{-2}): \quad 2A^{(2)}_{,u} + 2A^{(0)}(M + cc_{,u}) - BM - Bcc_{,u} - 2M_{,\theta} f^{(0)} = 0 \,, \tag{30}$$

$$\mathcal{L}_\eta g_{01} = 0 \quad (r^{-3}): \quad -2A^{(2)} + 2g^{(1)}(d_{,\theta} + 2d\cot\theta) - 2f^{(0)}[N + 2c(c_{,\theta} + c\cot\theta) + d(3d_{,\theta} + 4d\cot\theta)]$$
$$+ B[-N_{,\theta} - N\cot\theta + \epsilon^2 + \mu^2 - (d_{,\theta} + 2d\cot\theta)^2] = 0 \,, \tag{31}$$

$$\mathcal{L}_\eta g_{02} = 0 \quad (r^{-1}): \quad 2(A^{(1)}_{,\theta} - B_{,\theta} M + BM_{,\theta} - f^{(2)}_{,u} + f^{(0)} cc_{,u}) = 0 \,, \tag{32}$$

$$\mathcal{L}_\eta g_{03} = 0 \quad (r^{-1}): \quad f^{(0)}[4dc_{,u} - \cot\theta(d_{,\theta} + 2d\cot\theta) - (d_{,\theta} + 2d\cot\theta)_{,\theta}]$$
$$+ 2Bc_{,u}(d_{,\theta} + 2d\cot\theta) - g^{(2)}_{,u} - g^{(1)} c_{,u} = 0 \,, \tag{33}$$

$$\mathcal{L}_\eta g_{12} = 0 \quad (r^{-2}): \quad 3f^{(2)} + f^{(0)}(-3d^2 - \tfrac{1}{2}c^2) + 4g^{(1)}d - \tfrac{1}{2}B_{,\theta}(c^2 + d^2) - 4Bd(d_{,\theta} + 2d\cot\theta)$$
$$+ B[6N + 9(cc_{,\theta} + dd_{,\theta}) + 12(c^2 + d^2)\cot\theta] = 0 \,, \tag{34}$$

$$\mathcal{L}_\eta g_{13} = 0 \quad (r^{-2}): \; 3g^{(2)} - g^{(1)}c - 2f^{(0)}cd + 4Bc(d_{,\theta} + 2d\cot\theta) + 6B(P + c_{,\theta}\, d - cd_{,\theta}) = 0 , \tag{35}$$

$$\mathcal{L}_\eta g_{22} = 0 \quad (r^{-1}): \; 2A^{(2)} + 2A^{(1)}c + A^{(0)}(c^2 + d^2) + 2f^{(2)}{}_{,\theta}$$
$$+ f^{(0)}{}_{,\theta}(c^2 - 2d^2) + 8f^{(0)}d^2\cot\theta + 4dg^{(1)}{}_{,\theta} - 4g^{(1)}d\cot\theta$$
$$+ B[2C_{,u} + 5N_{,\theta} + N\cot\theta + 2cM - \epsilon^2 - \mu^2$$
$$+ (3c^2 + d^2)c_{,u} + 2cc_{,\theta\theta} + 3c_{,\theta}{}^2 - \tfrac{1}{2}c^2 - 4c^2\cot^2\theta$$
$$+ 2dd_{,\theta\theta} + 5d_{,\theta}{}^2 + 8dd_{,\theta}\cot\theta + 4d^2\cot^2\theta - \tfrac{1}{2}d^2] = 0 , \tag{36}$$

$$\mathcal{L}_\eta g_{23} = 0 \quad (r^{-1}): \; 2\{g^{(2)}{}_{,\theta} - g^{(2)}\cot\theta - g^{(1)}{}_{,\theta}c + g^{(1)}(c_{,\theta} + c\cot\theta) + 2dA^{(1)} - 4f^{(0)}d(c_{,\theta} + c\cot\theta)$$
$$+ B[2Md + 2P_{,\theta} - 2P\cot\theta + 2H_{,u} + 2dcc_{,u} - 2c_{,\theta}(d_{,\theta} + 4d\cot\theta) - 8dc\cot^2\theta]\} = 0 , \tag{37}$$

$$\mathcal{L}_\eta g_{33} = 0 \quad (r^{-1}): \; -2A^{(2)} + 2A^{(1)}c - A^{(0)}(c^2 + d^2) - 2f^{(2)}\cot\theta$$
$$+ f^{(0)}[-2d(2d_{,\theta} + 3d\cot\theta) - 3c(2c_{,\theta} + 3c\cot\theta)]$$
$$+ B[2C_{,u} - N_{,\theta} - 5N\cot\theta + 2Mc + \epsilon^2 + \mu^2$$
$$+ c_{,u}(3c^2 + d^2) - c_{,\theta}{}^2 - 10cc_{,\theta}\cot\theta - \tfrac{25}{2}c^2\cot^2\theta$$
$$+ \tfrac{1}{2}c^2\sin^{-2}\theta + d_{,\theta}{}^2 - 2dd_{,\theta}\cot\theta - \tfrac{9}{2}d^2\cot^2\theta + \tfrac{1}{2}d^2\sin^{-2}\theta] = 0 . \tag{38}$$

From Eq. (33) we find

$$g^{(2)} = [B_{,\theta}(d_{,\theta} + 2d\cot\theta)_{,\theta} + (d_{,\theta} + 2d\cot\theta)(B_{,\theta}\cot\theta + Bc_{,u}) - 3B_{,\theta}c_{,u}d]u + \tilde{g}^{(2)}(\theta) , \tag{39}$$

with $\tilde{g}^{(2)}(\theta)$ being an integration function. Eqs. (32) and (30) can also be solved easily,

$$f^{(2)} = 2u^2 Bc_{,u}^2\cot\theta + 2A^{(1)}B_{,\theta}B^{-1}u + \tilde{f}^{(2)}(\theta) , \tag{40}$$
$$A^{(2)} = -2u^2 B_{,\theta}c_{,u}^2\cot\theta + uB^{-2}\{A^{(1)}{}_{,\theta}B_{,\theta}B + A^{(1)}[B^2 c_{,u} + B_{,\theta}(-B_{,\theta} + B\cot\theta)]\} + a^{(2)}(\theta) , \tag{41}$$

with integration functions $\tilde{f}^{(2)}(\theta)$ and $a^{(2)}(\theta)$. From Eqs. (34) and (35) functions $N$ and $P$ can be calculated as

$$N = -A^{(1)}B_{,\theta}B^{-2}u + \tfrac{1}{2}B^{-1}[-\tilde{f}^{(2)}(\theta) + \tfrac{1}{2}B_{,\theta}d^2 - Bd(3d_{,\theta} + 4d\cot\theta)] , \tag{42}$$
$$P = \tfrac{1}{2}uB_{,\theta}B^{-1}[-(d_{,\theta} + 2d\cot\theta)_{,\theta} - (d_{,\theta} + 2d\cot\theta)\cot\theta + 2dc_{,u}] - \tfrac{1}{2}B^{-1}\tilde{g}^{(2)}(\theta) , \tag{43}$$

and Eq. (37) gives

$$H = \tfrac{1}{4}u^2 B_{,\theta}^2 B^{-2}[-(d_{,\theta} + 2d\cot\theta)_{,\theta} - (d_{,\theta} + 2d\cot\theta)\cot\theta + 2dc_{,u}] - \tfrac{1}{2}\tilde{g}^{(2)}(\theta)B_{,\theta}B^{-2}u + \tilde{H}(\theta) , \tag{44}$$

where $\tilde{H}(\theta)$ is an integration function. The addition and subtraction of Eqs. (36) and (38)/$\sin^2\theta$ yield

$$C = -\tfrac{1}{2}u^2 A^{(1)}B_{,\theta}^2 B^{-3} - \tfrac{1}{4}uB_{,\theta}B^{-2}[2\tilde{f}^{(2)}(\theta) + 6Bd(d_{,\theta} + 2d\cot\theta) - d^2(B_{,\theta} + 4B\cot\theta)] + \tilde{C}(\theta) , \tag{45}$$

with $\tilde{C}(\theta)$ an integration function, and

$$A^{(1)} = a_1 B^{-2} , \quad a_1 = \text{const} . \tag{46}$$

In addition we obtain the equation identical to Eq. (31), the solution of which is

$$a^{(2)} = \tfrac{1}{2}(\epsilon_0^2 + \mu_0^2)B^{-3} + \tfrac{1}{4}[\tilde{f}^{(2)}{}_{,\theta} + \tilde{f}^{(2)}(-3B_{,\theta}B^{-1} + \cot\theta)] + \tfrac{3}{4}dB(d_{,\theta} + 2d\cot\theta)_{,\theta}$$
$$+ \tfrac{1}{4}(d_{,\theta} + 2d\cot\theta)[5B(d_{,\theta} + 2d\cot\theta) + d(B_{,\theta} - 7B\cot\theta)]$$
$$+ \tfrac{1}{8}d^2[-2Bc_{,u} + 3B_{,\theta}B^{-1}(B_{,\theta} - 2B\cot\theta) + 4B(3\cot^2\theta + \sin^{-2}\theta)] . \tag{47}$$

Next we shall consider further consequences of the Einstein-Maxwell equations. From Eqs. (14), (22), (43), (44) we get the equations for $\tilde{g}^{(2)}$ and $d$,

$$\tilde{g}^{(2)}{}_{,\theta}B^2 - \tilde{g}^{(2)}B(-3B_{,\theta} + B\cot\theta) - 4da_1 = 0 , \tag{48}$$
$$d_{,\theta\theta\theta}B^2 + d_{,\theta\theta}B(-B_{,\theta} + 3B\cot\theta) + d_{,\theta}B[-B_{,\theta\theta} + 12B_{,\theta}\cot\theta + B(2 - 5\sin^{-2}\theta)]$$
$$+ d[B_{,\theta}B_{,\theta\theta} + 2BB_{,\theta\theta}\cot\theta + BB_{,\theta}(-12 + 4\sin^{-2}\theta) + B_{,\theta}^2\cot\theta + B^2\cot\theta(8 - 4\sin^{-2}\theta)] = 0 . \tag{49}$$

Eq. (49) is the third-order ordinary linear differential equation for $d(\theta)$. Another independent third-order equation for $d(\theta)$ follows from Eq. (6) in which we use (22) and (43). Now it can be shown that as a consequence of these two equations we find that $d(\theta)$ has to vanish,

$$d = 0 \ . \tag{50}$$

(The derivation proceeds as follows: first exclude $d_{,\theta\theta\theta}$, differentiate the resulting equation and subtract the result from one of the original equations; proceed similarly with the pair of the 2-nd order equations and get finally $d = 0$.)
It is then easy to see that Eq. (48) yields

$$\tilde{g}^{(2)} = g_0 B^{-3} \sin\theta \ , \quad g_0 = \text{const} \ . \tag{51}$$

Similarly, as a consequence of Eqs. (13), (42) and (45) we get

$$\tilde{f}^{(2)} = f_0 B^{-3} \sin\theta \ , \quad f_0 = \text{const} \ . \tag{52}$$

Eq. (5) is satisfied identically. Using (50) and (52) in (47) we arrive at

$$a^{(2)} = -\tfrac{1}{2} f_0 \sin\theta (3B_{,\theta} - B\cot\theta) B^{-4} + \tfrac{1}{2}(\epsilon_0^2 + \mu_0^2) B^{-3} \ . \tag{53}$$

Now we could write down all the results above explicitly in terms of parameters $a_0$, $b_0$ (characterizing the translations) by substituting for $B$ from Eqs. (26)–(28). Under the presence of the string the expressions are quite involved. Hence, we summarize them only in the case *without the string* ($\mathcal{C} = 1$). For the functions giving the Killing vector, Eqs. (22), (39)–(41), (46), (50)–(53) lead to

$$\begin{aligned}
A^{(0)} &= \tfrac{1}{2}(a_0 \cos\theta + b_0) \ , \quad A^{(1)} = \frac{a_1}{(-a_0\cos\theta + b_0)^2} \ , \\
A^{(2)} &= -u a_1 a_0 \frac{a_0(2\sin^2\theta + 1) + b_0 \cos\theta}{(-a_0\cos\theta + b_0)^4} + \tfrac{1}{2} f_0 \frac{-a_0(2\sin^2\theta + 1) + b_0 \cos\theta}{(-a_0\cos\theta + b_0)^4} + \frac{\epsilon_0^2 + \mu_0^2}{2(-a_0\cos\theta + b_0)^3} \ , \\
f^{(0)} &= -a_0 \sin\theta \ , \quad f^{(2)} = \frac{(u 2 a_1 a_0 + f_0)\sin\theta}{(-a_0\cos\theta + b_0)^3} \ , \\
g^{(2)} &= \frac{g_0 \sin\theta}{(-a_0\cos\theta + b_0)^3} \ .
\end{aligned} \tag{54}$$

Functions determining the asymptotic forms of the metric and electromagnetic field become

$$\begin{aligned}
c &= 0 \ , \quad d = 0 \ , \quad M = -\frac{a_1}{(-a_0\cos\theta + b_0)^3} \ , \\
N &= -\frac{u 2 a_1 a_0 + f_0}{2(-a_0\cos\theta + b_0)^4} \sin\theta \ , \quad P = -\frac{g_0 \sin\theta}{2(-a_0\cos\theta + b_0)^4} \ , \\
C &= -\frac{u a_1 a_0 + f_0}{2(-a_0\cos\theta + b_0)^5} u a_0 \sin^2\theta + \tilde{C}(\theta) \ , \quad H = -\frac{u g_0 a_0 \sin^2\theta}{2(-a_0\cos\theta + b_0)^5} + \tilde{H}(\theta) \ , \\
X &= 0 \ , \quad \epsilon = \frac{\epsilon_0}{(-a_0\cos\theta + b_0)^2} \ , \quad e = -\frac{u\epsilon_0 a_0 \sin\theta}{(-a_0\cos\theta + b_0)^3} + e_1(\theta) \ , \\
Y &= 0 \ , \quad \mu = \frac{\mu_0}{(-a_0\cos\theta + b_0)^2} \ , \quad f = \frac{u\mu_0 a_0 \sin\theta}{(-a_0\cos\theta + b_0)^3} + f_1(\theta) \ .
\end{aligned} \tag{55}$$

In Appendix A the complete asymptotic forms of the metric and electromagnetic field are written down in both the cases when the string is located along the axis and the cases without the string; then the results simplify considerably.

Here we give the explicit expansions of the translational Killing field, writing first the general expressions valid also under the presence of the string ($\mathcal{C} \neq 1$), in which case for functions $B$, $c$, $d$ expressions from (22), (26)–(28) and (50) have to be substituted into (24); then, after the arrows, we specialize them to the cases without the string ($\mathcal{C} = 1$). The contravariant components read

$$\begin{aligned}
\eta^u &= B \ \rightarrow \ -a_0\cos\theta + b_0 \ , \\
\eta^r &= B c_{,u} + B_{,\theta}\cot\theta - \frac{u c_{,u}^2 B}{r} + [-\tfrac{1}{2} u^2 c_{,u}^2 (3B_{,\theta}\cot\theta - Bc_{,u}) - (u a_1 B_{,\theta} + \tfrac{1}{2} f_0 \sin\theta) B_{,\theta} B^{-4}] \frac{1}{r^2} + O(r^{-3}) \\
&\rightarrow a_0 \cos\theta - \frac{(u 2 a_1 a_0 + f_0) a_0 \sin^2\theta}{2(-a_0\cos\theta + b_0)^4} \frac{1}{r^2} + O(r^{-3}) \ , \\
\eta^\theta &= -\frac{B_{,\theta}}{r} + \frac{u c_{,u} B_{,\theta}}{r^2} - \frac{u^2 c_{,u}^2 B_{,\theta}}{2r^3} + O(r^{-4}) \ \rightarrow \ -\frac{a_0 \sin\theta}{r} + O(r^{-4}) \ , \\
\eta^\phi &= O(r^{-4}) \ ,
\end{aligned} \tag{56}$$

where $B$ is given by (26)–(28), $c_{,u}$ by (22), $a_0$, $b_0$, $a_1$, $f_0$ are constants.

# IV. AXISYMMETRIC ELECTROVACUUM SPACETIMES WITH AN ASYMPTOTICALLY TRANSLATIONAL KILLING VECTOR AT NULL INFINITY

Since we wish to see how the character of the translational Killing vector is related to the properties of the spacetime we shall now first find its norm. Using the metric given in Appendix A we obtain the covariant components of the Killing vector:

$$\begin{aligned}
\eta_u &= Bc_{,u} + B_{,\theta}\cot\theta + B + (uc_{,u}^2 B + 2a_1 B^{-2})\frac{1}{r} \\
&\quad + [ua_1 B^{-4}(2c_{,u} B^2 - 3B_{,\theta}^2 + 2BB_{,\theta}\cot\theta) + \tfrac{1}{2}f_0\sin\theta\, B^{-4}(-3B_{,\theta} + 2B\cot\theta) + (\epsilon_0^2+\mu_0^2)B^{-3}]\frac{1}{r^2} + O(r^{-3}) \\
&\to b_0 + \frac{2a_1}{(-a_0\cos\theta+b_0)^2}\frac{1}{r} + \left[(u2a_1 a_0 + f_0)\frac{-a_0(2+\sin^2\theta)+2b_0\cos\theta}{(-a_0\cos\theta+b_0)^4} + \frac{\epsilon_0^2+\mu_0^2}{(-a_0\cos\theta+b_0)^3}\right]\frac{1}{r^2} + O(r^{-3})\,, \\
\eta_r &= B - \frac{u^2 c_{,u}^2 B}{2r^2} + O(r^{-3}) \quad \to \quad -a_0\cos\theta + b_0 + O(r^{-3})\,, \qquad (57)\\
\eta_\theta &= B_{,\theta}\, r + uc_{,u} B_{,\theta} + [\tfrac{1}{2}u^2 c_{,u}^2 (B_{,\theta} - 4B\cot\theta) - (u2a_1 B_{,\theta} + f_0\sin\theta)B^{-3}]\frac{1}{r} + O(r^{-2})\\
&\to a_0\sin\theta\, r - \frac{(u2a_1 a_0 + f_0)\sin\theta}{(-a_0\cos\theta+b_0)^3}\frac{1}{r} + O(r^{-2})\,,\\
\eta_\phi &= -\frac{g_0\sin^2\theta}{B^3}\frac{1}{r} + O(r^{-2}) \quad \to \quad -\frac{g_0\sin^2\theta}{(-a_0\cos\theta+b_0)^3}\frac{1}{r} + O(r^{-2})\,.
\end{aligned}$$

Combining these with contravariant components (56) we find the square of the norm of the Killing vector, $\|\eta\|^2 = g_{\alpha\beta}\eta^\alpha\eta^\beta$, to have asymptotically the form

$$\begin{aligned}
\|\eta\|^2 &= (b_0^2 - a_0^2) + \frac{2a_1}{B}\frac{1}{r} + [u2a_1 B^{-3}(B^2 c_{,u} - B_{,\theta}^2 + BB_{,\theta}\cot\theta) \\
&\quad + f_0\sin\theta\, B^{-3}(-B_{,\theta} + B\cot\theta) + (\epsilon_0^2+\mu_0^2)B^{-2}]\frac{1}{r^2} + O(r^{-3}) \\
&\to (b_0^2 - a_0^2) + \frac{2a_1}{-a_0\cos\theta+b_0}\frac{1}{r} + \left[\frac{(u2a_1 a_0 + f_0)(-a_0+b_0\cos\theta)}{(-a_0\cos\theta+b_0)^3} + \frac{\epsilon_0^2+\mu_0^2}{(-a_0\cos\theta+b_0)^2}\right]\frac{1}{r^2} + O(r^{-3})\,. \quad (58)
\end{aligned}$$

Hence, we see that the Killing vector is asymptotically timelike if $b_0^2 > a_0^2$, null if $b_0^2 = a_0^2$, and spacelike if $b_0^2 < a_0^2$.

Now the resulting expressions for the metric and field given in the previous section enable us to formulate the following

### Theorem 1

*If an axisymmetric electrovacuum spacetime, with in general an infinite cosmic string along the axis $\theta = 0,\ \pi$, admits a local $\mathcal{I}^+$ and an asymptotically spacelike translational Killing vector, given by Eq. (56) with $a_0^2 > b_0^2$, and if the part of the Bondi mass aspect M which is not caused by the string is non-vanishing so that constant $a_1 \neq 0$ (see Eqs. (22), (26)–(28), (46) if there is the string, and the simple expression for M in Eq. (55) in the case without the string), or if $a_1 = 0$, but any of the other constants $f_0$, $g_0$ appearing in the metric functions N, P, H, C is non-vanishing (see Eqs. (42)–(45), (51), (52)), then $\mathcal{I}^+$ contains singular generators at $\theta = \theta_0 \neq 0,\ \pi$, where $\theta_0$ is given by Eq. (59), in addition to those at $\theta = 0,\ \pi$ due to the string. If the translational Killing vector is null, $a_0^2 = b_0^2$, $\mathcal{I}^+$ is singular at $\theta = 0,\ \pi$. If it is timelike, $a_0^2 < b_0^2$, the only singular generators occur at $\theta = 0,\ \pi$, and are due to the presence of the string.*

In particular, if no string is present along the axis, we can formulate

### Theorem 2

*If an axisymmetric electrovacuum spacetime with a non-vanishing Bondi mass m (defined by Eq. (60) with the mass aspect M given in Eq. (55)) admits an asymptotically translational Killing vector and a complete cross section of $\mathcal{I}^+$, then the translational Killing vector is timelike and spacetime is thus stationary.*

*Proof*

From Eq. (58) we see that if $a_0^2 > b_0^2$ the translational Killing vector is spacelike. Then there exists $\theta = \theta_0$ given by the relation

$$\cos\theta_0 = \frac{(a_0+b_0)^{2/\mathcal{C}} - (a_0^2-b_0^2)^{1/\mathcal{C}}}{(a_0+b_0)^{2/\mathcal{C}} + (a_0^2-b_0^2)^{1/\mathcal{C}}} \ , \tag{59}$$

for which the function $B(\theta=\theta_0)$ given by (26)–(28) vanishes and, therefore, the mass aspect $M$ (given in (22) and (46) with $a_1 \neq 0$), and in general (with $f_0 \neq 0$, $g_0 \neq 0$) other metric and field functions $N$, $P$, $H$, $C$ in (42)–(45) diverge; consequently, the metric components given in Appendix A diverge as well. This case thus corresponds to cylindrical waves for which $\mathcal{I}^+$ does not exist for $\theta = \theta_0$. In particular, if $b_0 = 0$, Eq. (59) implies that $\theta_0 = \pi/2$. In Appendix C a special class of Einstein-Rosen waves is considered for which $a_1 = 0$, however, $\mathcal{I}^+$ is singular at $\theta_0 = \pi/2$ due to $f_0 \neq 0$. If the translational Killing vector is null, $|a_0| = |b_0|$, then $\theta_0 = 0$ or $\pi$ – which corresponds to a wave propagating along the symmetry axis. However, even if the translational Killing vector is spacelike or null, $\mathcal{I}^+$ may be regular if in its neighbourhood the spacetime is flat. Then the mass aspect $M$, and hence the Bondi mass $m$, as well as functions $A^{(1)}$, $N$, $P$, etc will vanish and no singularity arises for $B(\theta=\theta_0) = 0$.

The presence of the string implies that $\mathcal{I}^+$ is singular at $\theta = 0$, $\pi$. If no string is present and $\mathcal{I}^+$ admits a complete cross section, and if the Bondi mass is non-vanishing (so that necessarily $a_1 \neq 0$), then the translational Killing vector must be timelike, $b_0^2 > a_0^2$, so that $B = -a_0\cos\theta + b_0$ does not vanish for any $\theta$ and the mass aspect given in Eq. (55) is regular everywhere.

The total mass in the case without the string is given by

$$m = \tfrac{1}{2}\int_0^\pi M(u,\ \theta)\sin\theta\ \mathrm{d}\theta = -\frac{a_1 b_0}{(b_0^2-a_0^2)^2} \ , \tag{60}$$

which is finite for $b_0^2 \neq a_0^2$. Taking $b_0 > 0$, then for physical, stationary systems the constant parameter $a_1 < 0$ so that $m > 0$. This is explicitly seen in Appendix B where the case of the Schwarzschild metric (with in general a cosmic string) is discussed. Clearly, in the case of the timelike Killing vector, the factors $-a_0\cos\theta + b_0$ appearing in Eq. (55) correspond, using the terminology of [10], to the "Doppler shift of the mass aspect" and other Bondi's functions which occur when the system is boosted with respect to Bondi's frame with the boost parameter $-\nu$ so that its velocity is $v = -\tanh\nu$. Writing

$$b_0 = \lambda\cosh\nu\ , \quad a_0 = \lambda\sinh\nu\ , \quad \lambda = \mathrm{const}\ > 0\ , \tag{61}$$

and putting

$$a_1 = -\bar{m}\lambda^3\ , \quad \bar{m} = \mathrm{const}\ , \tag{62}$$

we get the mass aspect in Eq. (55) in the form $M = \bar{m}/(\cosh\nu - \sinh\nu\cos\theta)^3$, which exactly corresponds to the formula (see Eq. (72) in [10]) for the Schwarzschild mass $\bar{m}$ moving along the axis of symmetry with constant velocity $-\tanh\nu$. Using (61) and (62) in our expression (60) for the Bondi mass we obtain

$$m = \bar{m}\cosh\nu\ , \tag{63}$$

as expected. The calculation of the Bondi momentum along the axis (see e.g. App. D in [11]) yields

$$P^z = \tfrac{1}{2}\int_0^\pi M(u,\ \theta)\sin\theta\cos\theta\ \mathrm{d}\theta = -\frac{a_1 a_0}{(b_0^2-a_0^2)^2} = \bar{m}\sinh\nu\ . \tag{64}$$

### ACKNOWLEDGMENTS

We are grateful to Piotr Chruściel who, after reading our paper [1], suggested that it would be of interest to analyze the translational case in a greater detail; J. B. enjoyed several discussions with Piotr Chruściel in the Albert-Einstein Institute in Potsdam; P. C. made also helpful suggestions in the presentation of this paper. A. P. thanks for the kind hospitality of the Institute of Theoretical Physics of the F. Schiller-University in Jena where part of this work was done, and Vojtěch Pravda for discussions. J. B. thanks Swiss Nationalfonds for support and Petr Hájíček for discussions and kind hospitality at the Institute of Theoretical Physics in Berne where this work was finished. We both acknowledge the support from Grant No. GACR-202/99/0261 of the Czech Republic and A. P. also the support from the Grant No. GACR-201/97/0217.

# APPENDIX A: THE ASYMPTOTIC FORM OF THE METRIC AND ELECTROMAGNETIC FIELD

Here we give the asymptotic expansions at null infinity of all metric and electromagnetic field components for axisymmetric spacetimes with a translational Killing vector by substituting the appropriate functions entering the metric and the field from Section III. In all formulas below we first write the expressions in which functions $B$, $c_{,u}$ etc. are given by (22), (27), (28), (42)–(46), (50), corresponding to the cases in which there is also the string along the axis ($\mathcal{C} \neq 1$). The simpler expressions after the arrows are explicit results for the cases without the string ($\mathcal{C} = 1$):

$$g_{uu} = 1 + 2(uc_{,u}^2 + a_1 B^{-3})\frac{1}{r}$$
$$+ \left[u2a_1 B^{-5}(B^2 c_{,u} - 2B_{,\theta}^2 + BB_{,\theta}\cot\theta) - f_0\sin\theta B^{-5}(2B_{,\theta} - B\cot\theta) + B^{-4}(\epsilon_0^2 + \mu_0^2)\right]\frac{1}{r^2}$$
$$+ \Big\{ -u^3 c_{,u}^4 - \tfrac{1}{2}u^2 a_1 B^{-7}\big[B^2 c_{,u}(-2B^2 c_{,u} + 23B_{,\theta}^2 - 8BB_{,\theta}\cot\theta)$$
$$+ B_{,\theta}^2\left(-15B_{,\theta}^2 + 20BB_{,\theta}\cot\theta + 2B^2(1-2\cot^2\theta)\right)\big]$$
$$+ \tfrac{1}{2}uf_0\sin\theta B^{-7}\big[2B^2 c_{,u}(-7B_{,\theta} + 2B\cot\theta) - B_{,\theta}\left(-15B_{,\theta}^2 + 20BB_{,\theta}\cot\theta + 2B^2(1-2\cot^2\theta)\right)\big]$$
$$- u(\epsilon_0^2 + \mu_0^2)B^{-6}(-2B^2 c_{,u} + 3B_{,\theta}^2 - 2BB_{,\theta}\cot\theta) - \tfrac{1}{2}\tilde{C}_{\theta\theta} - \tfrac{3}{2}\tilde{C}_{,\theta}\cot\theta + \tilde{C}$$
$$+ B^{-2}[\mu_0(f_{1,\theta} + f_1\cot\theta) - \epsilon_0(e_{1,\theta} + e_1\cot\theta)]\Big\}\frac{1}{r^3} + O(r^{-4}) ,$$

$$\to \; 1 + \frac{2a_1}{(-a_0\cos\theta + b_0)^3}\frac{1}{r} + \Big\{ -\frac{u2a_1 a_0 + f_0}{(-a_0\cos\theta + b_0)^5}[a_0(\sin^2\theta + 1) - b_0\cos\theta] + \frac{\epsilon_0^2 + \mu_0^2}{(-a_0\cos\theta + b_0)^4}\Big\}\frac{1}{r^2}$$
$$+ \Big\{ -ua_0 \frac{ua_1 a_0 + f_0}{(-a_0\cos\theta + b_0)^7}[a_0^2(-15 + 12\cos^2\theta - \cos^4\theta) + 8a_0 b_0\cos\theta(1+\sin^2\theta)$$
$$+ 2b_0^2(-2 + 3\sin^2\theta)] - ua_0\frac{\epsilon_0^2 + \mu_0^2}{(-a_0\cos\theta + b_0)^6}[a_0(2+\sin^2\theta) - 2b_0\cos\theta] - \tfrac{1}{2}\tilde{C}_{\theta\theta} - \tfrac{3}{2}\tilde{C}_{,\theta}\cot\theta + \tilde{C}$$
$$+ \frac{1}{(-a_0\cos\theta + b_0)^2}[\mu_0(f_{1,\theta} + f_1\cot\theta) - \epsilon_0(e_{1,\theta} + e_1\cot\theta)]\Big\}\frac{1}{r^3} + O(r^{-4}) ,$$

$$g_{ur} = 1 - \frac{u^2 c_{,u}^2}{2r^2} + O(r^{-3}) \quad \to \; 1 + O(r^{-3}) ,$$

$$g_{u\theta} = (-2u^2 c_{,u}^2\cot\theta - 2ua_1 B_{,\theta} B^{-4} - f_0\sin\theta B^{-4})\frac{1}{r}$$
$$+ \Big[-\tfrac{1}{4}u^2 a_1 B_{,\theta} B^{-6}(16B^2 c_{,u} - 15B_{,\theta}^2 + 12BB_{,\theta}\cot\theta) + \tfrac{1}{4}uf_0\sin\theta B^{-6}(-8B^2 c_{,u} + 15B_{,\theta}^2 - 12BB_{,\theta}\cot\theta)$$
$$- uB_{,\theta} B^{-5}(\epsilon_0^2 + \mu_0^2) + \tfrac{3}{2}\tilde{C}_{,\theta} + 3\tilde{C}\cot\theta + (\epsilon_0 e_1 + \mu_0 f_1)B^{-2}\Big]\frac{1}{r^2} + O(r^{-3})$$

$$\to \; -\frac{2ua_1 a_0 + f_0}{(-a_0\cos\theta + b_0)^4}\frac{\sin\theta}{r} + \Big\{ 3ua_0\sin\theta\frac{ua_1 a_0 + f_0}{(-a_0\cos\theta + b_0)^6}[a_0(4+\sin^2\theta) - 4b_0\cos\theta]$$
$$- ua_0\sin\theta\frac{\epsilon_0^2 + \mu_0^2}{(-a_0\cos\theta + b_0)^5} + \tfrac{3}{2}\tilde{C}_{,\theta} + 3\tilde{C}\cot\theta + \frac{\epsilon_0 e_1 + \mu_0 f_1}{(-a_0\cos\theta + b_0)^2}\Big\}\frac{1}{r^2} + O(r^{-3}) ,$$

$$g_{u\phi} = -\frac{g_0\sin^2\theta}{B^4}\frac{1}{r} + \Big[\tfrac{1}{4}ug_0\sin\theta\, B^{-6}(-4c_{,u} B^2 + 15B_{,\theta}^2 - 12BB_{,\theta}\cot\theta)$$
$$+ \tfrac{3}{2}\tilde{H}_{,\theta} + 3\tilde{H}\cot\theta + (\epsilon_0 f_1 + \mu_0 e_1)B^{-2}\Big]\frac{\sin\theta}{r^2} + O(r^{-3})$$

$$\to \; -\frac{g_0\sin^2\theta}{(-a_0\cos\theta + b_0)^4}\frac{1}{r} + \Big\{ \frac{3ug_0 a_0\sin\theta}{4(-a_0\cos\theta + b_0)^6}[a_0(4+\sin^2\theta) - 4b_0\cos\theta]$$
$$+ \tfrac{3}{2}\tilde{H}_{,\theta} + 3\tilde{H}\cot\theta + \frac{\epsilon_0 f_1 + \mu_0 e_1}{(-a_0\cos\theta + b_0)^2}\Big\}\frac{\sin\theta}{r^2} + O(r^{-3}) ,$$

$$g_{rr} = 0 ,$$
$$g_{r\theta} = 0 ,$$
$$g_{r\phi} = 0 ,$$
$$g_{\theta\theta} = -r^2 - 2uc_{,u}\, r - 2u^2 c_{,u}^2 + (-u^3 c_{,u}^3 + u^2 a_1 B_{,\theta}^2 B^{-5} + uf_0 B_{,\theta}\sin\theta B^{-5} - 2\tilde{C})\frac{1}{r} + O(r^{-2})$$

$$\rightarrow -r^2 + \left[ua_0 \sin^2\theta \frac{ua_1 a_0 + f_0}{(-a_0 \cos\theta + b_0)^5} - 2\tilde{C}\right]\frac{1}{r} + O(r^{-2}) \;,$$

$$g_{\theta\phi} = -(-ug_0 B_{,\theta} \sin\theta \; B^{-5} + 2\tilde{H})\frac{\sin\theta}{r} + O(r^{-2}) \quad \rightarrow \quad -\left[-\frac{ug_0 a_0 \sin^2\theta}{(-a_0 \cos\theta + b_0)^5} + 2\tilde{H}\right]\frac{\sin\theta}{r} + O(r^{-2}) \;,$$

$$g_{\phi\phi} = -\sin^2\theta \; r^2 + 2uc_{,u} \sin^2\theta \; r - 2u^2 c_{,u}^2 \sin^2\theta + (u^3 c_{,u}^3 - u^2 a_1 B_{,\theta}^2 \; B^{-5} - uf_0 B_{,\theta} \sin\theta B^{-5} + 2\tilde{C})\frac{\sin^2\theta}{r} + O(r^{-2})$$

$$\rightarrow -\sin^2\theta \; r^2 + \left[-ua_0 \sin^2\theta \frac{ua_1 a_0 + f_0}{(-a_0 \cos\theta + b_0)^5} + 2\tilde{C}\right]\frac{\sin^2\theta}{r} + O(r^{-2}) \;,$$

$$F_{01} = -\epsilon_0 B^{-2}\frac{1}{r^2} + \left[u\epsilon_0 B^{-4}(-BB_{,\theta\theta} + 3B_{,\theta}^2 - BB_{,\theta}\cot\theta) + e_{1,\theta} + e_1 \cot\theta\right]\frac{1}{r^3} + O(r^{-4})$$

$$\rightarrow -\frac{\epsilon_0}{(-a_0 \cos\theta + b_0)^2}\frac{1}{r^2} + \left[\frac{u\epsilon_0 a_0}{(-a_0 \cos\theta + b_0)^4}(a_0 + a_0 \sin^2\theta - 2b_0 \cos\theta) + e_{1,\theta} + e_1 \cot\theta\right]\frac{1}{r^3} + O(r^{-4}) \;,$$

$$F_{02} = -\epsilon_0 B_{,\theta} B^{-3}\frac{1}{r} + O(r^{-2}) \quad \rightarrow \quad -\frac{\epsilon_0 a_0 \sin\theta}{(-a_0 \cos\theta + b_0)^3}\frac{1}{r} + O(r^{-2}) \;,$$

$$F_{03} = -\mu_0 B_{,\theta} B^{-3} \sin\theta\frac{1}{r} + O(r^{-2}) \quad \rightarrow \quad -\frac{\mu_0 a_0 \sin\theta}{(-a_0 \cos\theta + b_0)^3}\frac{\sin\theta}{r} + O(r^{-2}) \;,$$

$$F_{12} = (-\epsilon_0 B_{,\theta} B^{-3} u + e_1)\frac{1}{r^2} + O(r^{-3}) \quad \rightarrow \quad \left[-\frac{u\epsilon_0 a_0 \sin\theta}{(-a_0 \cos\theta + b_0)^3} + e_1\right]\frac{1}{r^2} + O(r^{-3}) \;,$$

$$F_{13} = (\mu_0 B_{,\theta} B^{-3} u + f_1) \sin\theta\frac{1}{r^2} + O(r^{-3}) \quad \rightarrow \quad \left[\frac{u\mu_0 a_0 \sin\theta}{(-a_0 \cos\theta + b_0)^3} + f_1\right]\frac{\sin\theta}{r^2} + O(r^{-3}) \;,$$

$$F_{23} = -\mu_0 B^{-2} \sin\theta - [\mu_0 B^{-4}(-3B_{,\theta}^2 + BB_{,\theta}\cot\theta + BB_{,\theta\theta})u + f_{1,\theta} + f_1 \cot\theta]\frac{\sin\theta}{r} + O(r^{-2})$$

$$\rightarrow -\frac{\mu_0}{(-a_0 \cos\theta + b_0)^2}\sin\theta + \left[\frac{u\mu_0 a_0}{(-a_0 \cos\theta + b_0)^4}(a_0 + a_0 \sin^2\theta - 2b_0 \cos\theta) - f_{1,\theta} - f_1 \cot\theta\right]\frac{\sin\theta}{r^3} + O(r^{-4}) \;.$$

## APPENDIX B: EXAMPLES

### 1. Schwarzschild black hole with a cosmic string

As an example of an axisymmetric spacetime with a translational timelike Killing field and only local $\mathcal{I}^+$ which does not admit compact smooth cross sections we consider the Schwarzschild black hole with a cosmic string along the $z$-axis. In coordinates $\{\bar{t}, \bar{r}, \bar{\vartheta}, \bar{\phi}\}$ the metric is given by (see e.g. [9])

$$ds^2 = \left(1 - \frac{2\bar{m}}{\bar{r}}\right)d\bar{t}^2 - \left(1 - \frac{2\bar{m}}{\bar{r}}\right)^{-1}d\bar{r}^2 - \bar{r}^2(d\bar{\vartheta}^2 + \mathcal{C}^2 \sin^2\bar{\vartheta} \; d\bar{\phi}^2) \;. \tag{B1}$$

Going over to Weyl's coordinates $\{\bar{t}, \bar{\rho}, \bar{z}, \bar{\phi}\}$ by putting

$$\bar{r} = \bar{m} + \tfrac{1}{2}(\bar{r}_+ + \bar{r}_-) \;, \quad \cos\bar{\vartheta} = (\bar{r}_+ - \bar{r}_-)/2\bar{m} \;, \quad \bar{r}_\pm^2 = \bar{\rho}^2 + (\bar{z} \pm \bar{m})^2 \;, \tag{B2}$$

or inversly

$$\bar{\rho} = \sqrt{\bar{r}^2 - 2\bar{m}\bar{r}} \; \sin\bar{\vartheta} \;, \quad \bar{z} = (\bar{r} - \bar{m}) \; \cos\bar{\vartheta} \;, \tag{B3}$$

we find

$$ds^2 = e^{2\psi}d\bar{t}^2 - e^{-2\psi}[e^{2\sigma}(d\bar{\rho}^2 + d\bar{z}^2) + \bar{\rho}^2\mathcal{C}^2 d\bar{\phi}^2] \;, \tag{B4}$$

where the metric functions are given by

$$e^{2\psi} = 1 - \frac{2\bar{m}}{\bar{r}} = \frac{(\bar{r}_+ + \bar{r}_-)^2 - 4\bar{m}^2}{(\bar{r}_+ + \bar{r}_- + 2\bar{m})^2} \;,$$

$$e^{2\sigma} = \frac{(\bar{r}_+ + \bar{r}_-)^2 - 4\bar{m}^2}{4\bar{r}_+\bar{r}_-} \;. \tag{B5}$$

Transforming into spherical coordinates $\{R, \vartheta, \phi\}$ by $\rho = R \sin \vartheta$, $z = R \cos \vartheta$, $\phi = \phi$, and introducing retarded time $U = t - R$, we get

$$ds^2 = e^{2\psi}dU^2 + 2e^{2\psi}dU\,dR - (e^{2(\sigma-\psi)} - e^{2\psi})dR^2 - e^{2(\sigma-\psi)}R^2 d\vartheta^2 - R^2 \mathcal{C}^2 \sin^2 \vartheta\, e^{-2\psi}d\phi^2 \,. \tag{B6}$$

Expanding the functions $e^{2\psi}$ and $e^{2\sigma}$ in $R^{-k}$,

$$\begin{aligned}
e^{2\psi} &= 1 - \frac{2\bar{m}}{R} + \frac{2\bar{m}^2}{R^2} - \bar{m}^3 \frac{1+\cos^2\vartheta}{R^3} + \frac{2\bar{m}^4 \cos^2\vartheta}{R^4} + \cdots \,, \\
e^{-2\psi} &= 1 + \frac{2\bar{m}}{R} + \frac{2\bar{m}^2}{R^2} + \bar{m}^3 \frac{1+\cos^2\vartheta}{R^3} + \frac{2\bar{m}^4 \cos^2\vartheta}{R^4} + \cdots \,, \\
e^{2\sigma} &= 1 - \frac{\bar{m}^2 \sin^2\vartheta}{R^2} + \bar{m}^4 \frac{1 - 4\cos^2\vartheta + 3\cos^4\vartheta}{R^4} + \cdots \,,
\end{aligned} \tag{B7}$$

we find the asymptotic form of the metric to read

$$\begin{aligned}
ds^2 &= \left[1 - \frac{2\bar{m}}{R} + \frac{2\bar{m}^2}{R^2} - \bar{m}^3 \frac{1+\cos^2\vartheta}{R^3} + \cdots \right]dU^2 + 2\left[1 - \frac{2\bar{m}}{R} + \frac{2\bar{m}^2}{R^2} - \bar{m}^3 \frac{1+\cos^2\vartheta}{R^3} + \cdots \right]dUdR \\
&\quad - \left[\frac{4\bar{m}}{R} - \frac{\bar{m}^2 \sin^2\vartheta}{R^2} + \cdots \right]dR^2 - \left[R^2 + 2\bar{m}R + \bar{m}^2(1 + \cos^2\vartheta) + \cdots \right]d\vartheta^2 \\
&\quad - \left[R^2 + 2\bar{m}R + 2\bar{m}^2 \sin^2\vartheta + \cdots \right]\mathcal{C}^2 \sin^2\vartheta\,d\phi^2 \,.
\end{aligned} \tag{B8}$$

Let us assume the transformation to Bondi's coordinates $\{u, r, \theta, \phi\}$ can be written in the form

$$\begin{aligned}
U &= \overset{ln}{\pi}(u,\theta)\ln r + \overset{o}{\pi}(u,\theta) + \overset{1}{\pi}(u,\theta)/r + \cdots \,, \\
R &= q(u,\theta)r + \overset{o}{\sigma}(u,\theta) + \overset{1}{\sigma}(u,\theta)/r + \cdots \,, \\
\vartheta &= \overset{o}{\tau}(u,\theta) + \overset{1}{\tau}(u,\theta)/r + \overset{2}{\tau}(u,\theta)/r^2 + \cdots \,.
\end{aligned} \tag{B9}$$

Transforming the metric (B8) into Bondi's coordinates by (B9) and comparing it with the standard form of the Bondi metric we determine functions entering the transformation (B9) as follows:

$$\begin{aligned}
&\overset{ln}{\pi} = 2\bar{m} \,, \quad \overset{o}{\pi} = \frac{u}{q} + 2\bar{m}\ln q + \text{const} \,, \quad \overset{1}{\pi} = -\frac{q_{,\theta}^2}{q^3}\frac{u^2}{2} + \frac{\bar{m}}{q^2}(q_{,\theta}^2 + q^2 - 1)u - \frac{4\bar{m}^2}{q} + \frac{\bar{m}^2 \sin^2\theta}{2q^3\mathcal{C}^2} \,, \\
&q = \frac{\sin\theta}{2\mathcal{C}}\left[\left(\chi\frac{\sin\theta}{\cos\theta + 1}\right)^{\mathcal{C}} + \left(\chi\frac{\sin\theta}{\cos\theta + 1}\right)^{-\mathcal{C}}\right] \,, \quad \chi = \text{const} \,, \\
&\overset{o}{\sigma} = \tfrac{1}{2}uq^{-1}(q_{,\theta}^2 + q^2 - 1) - \bar{m} \,, \quad \overset{1}{\sigma} = \tfrac{1}{2}u^2\left(\frac{q_{,\theta}^2}{q^3} + qc_{,u}^2\right) - \frac{\bar{m}^2}{2\mathcal{C}^2 q^3}\sin^2\theta \,, \\
&\overset{o}{\tau}_{,u} = 0 \,, \quad \overset{o}{\tau}_{,\theta} = \pm\frac{1}{q} \,, \quad \overset{1}{\tau} = \pm\frac{q_{,\theta}\,u}{q^2} \,, \quad \overset{2}{\tau} = -\tfrac{1}{2}u^2 q_{,\theta}\,q^{-4}(q_{,\theta}^2 + q^2 - 1) + \frac{\bar{m}^2}{2\mathcal{C}^2 q^4}\sin^2\theta(q_{,\theta} - q\cot\theta) \,.
\end{aligned} \tag{B10}$$

We also find

$$c = u\frac{\mathcal{C}^2 - 1}{2\sin^2\theta} \,. \tag{B11}$$

Using now these functions (choosing the + sign and the simplest unboosted Bondi's system with $\chi = 1$) to transform the Killing vector generating translations along the $\bar{t}$-axis, which in the original Schwarzschild coordinates is given by

$$\eta^\alpha = [\,b_0,\ 0,\ 0,\ 0\,] \,, \tag{B12}$$

into Bondi's coordinates, we obtain

$$\begin{aligned}
\eta_u &= b_0 \left[\frac{1}{2q}(q_{,\theta}^2 + q^2 + 1) - \left(-uqc_{,u}^2 + \frac{2\bar{m}}{q^2}\right)\frac{1}{r} + \cdots \right] \,, \\
\eta_r &= b_0 \left(q - \frac{u^2 qc_{,u}^2}{2r^2} + \cdots \right) \\
\eta_\theta &= b_0 \left\{q_{,\theta}\,r + uq_{,\theta}\,c_{,u} + \left[\tfrac{1}{2}u^2 c_{,u}^2 (q_{,\theta} - 4q\cot) + \frac{2\bar{m}uq_{,\theta}}{q^3}\right]\frac{1}{r} + \cdots \right\} \,, \\
\eta_\phi &= 0 \,,
\end{aligned} \tag{B13}$$

where $q$ and $c$ are given in (B10) and (B11).

Comparing (B13) with the asymptotically translational timelike Killing vector (57), in which we put $a_0 = 0$, the results are in agreement if the constant parameters (cf. (61), (62) with $\nu = 0$)

$$a_1 = -\bar{m}b_0^3 = -\bar{m}\lambda^3 , \quad f_0 = g_0 = 0 . \tag{B14}$$

Let us notice yet that in the work of Bondi *et al* [10] the Weyl metric without the string is transformed into Bondi's coordinates, whereas in [4] the flat spacetime with an infinite string along the $z$-axis is converted to these coordinates. Our results (B9), (B10) thus combine those of Refs. [10] and [4].

### 2. Einstein-Rosen waves with a cosmic string

As an example of a spacetime with an axial and cylindrical symmetry we consider a class of Einstein-Rosen cylindrical waves with, in addition, a string along the $z$-axis. The metric is given by

$$ds^2 = e^{2(\gamma-\psi)}(dt^2 - d\rho^2) - e^{2\psi}dz^2 - \mathcal{C}^2\rho^2 e^{-2\psi}d\phi^2 , \tag{B15}$$

which, transforming to "spherical coordinates" $\{R, \vartheta, \phi\}$ by $\rho = R\sin\vartheta$, $z = R\cos\vartheta$, $\phi = \phi$, and introducing "retarded time" $U = t - R$, goes over into the form

$$ds^2 = e^{2(\gamma-\psi)}dU^2 + 2e^{2(\gamma-\psi)}dU\,dR + (e^{2(\gamma-\psi)} - e^{2\psi})\cos^2\vartheta\,dR^2 - (e^{2(\gamma-\psi)}\cos^2\vartheta + e^{2\psi}\sin^2\vartheta)R^2 d\vartheta^2$$
$$+ 2(d^{2\psi} - d^{2(\gamma-\psi)})R\sin\vartheta\,\cos\vartheta\,dRd\vartheta - R^2\mathcal{C}^2\sin^2\vartheta\,e^{-2\psi}d\phi^2 . \tag{B16}$$

We shall consider one class of the waves analyzed in detail in [7] – those representing the "averaged" time-symmetric solutions (the space average of the time derivative of $\psi$ vanishes at $t = 0$), which in particular occur in the time-symmetric case. It is proved in [7] that for the data of compact support, but also in more general cases as for the well-known Weber-Wheeler-Bonnor time-symmetric pulse, the functions $\psi$ and $\gamma$ in coordinates $\{U, R, \vartheta, \phi\}$ have expansions at large $R$ with $U, \vartheta, \phi$ fixed of the form

$$\psi = \frac{L}{\cos^3\vartheta}\left[-\frac{1}{R^2} + \frac{(2+\sin^2\vartheta)U}{\cos^2\vartheta}\frac{1}{R^3} + \cdots\right] , \tag{B17}$$

$$\gamma = \frac{L^2\sin^2\vartheta}{4\cos^8\vartheta}\left[(8+\sin^2\vartheta)\frac{1}{R^4} - \frac{24(2+\sin^2\vartheta)U}{\cos^2\vartheta}\frac{1}{R^5} + \cdots\right] , \tag{B18}$$

where $L = $ const. The metric (B16) then becomes

$$ds^2 = \left[1 + \frac{2L}{\cos^3\vartheta}\frac{1}{R^2} + \frac{2LU(-3+\cos^2\vartheta)}{\cos^5\vartheta}\frac{1}{R^3} + \cdots\right]dU^2 + 2\left[1 + \frac{2L}{\cos^3\vartheta}\frac{1}{R^2} + \frac{2LU(-3+\cos^2\vartheta)}{\cos^5\vartheta}\frac{1}{R^3} + \cdots\right]dU\,dR$$
$$+ \left[\frac{4L}{\cos\vartheta}\frac{1}{R^2} + \frac{4LU(-3+\cos^2\vartheta)}{\cos^3\vartheta}\frac{1}{R^3} + \cdots\right]dR^2 - \left[\frac{8L}{\cos^2\vartheta}\frac{1}{R} + \frac{8LU(-3+\cos^2\vartheta)}{\cos^4\vartheta}\frac{1}{R^2} + \cdots\right]\sin\vartheta\,dRd\vartheta$$
$$- \left[1 + \frac{2L(\cos^2\vartheta - \sin^2\vartheta)}{\cos^3\vartheta}\frac{1}{R^2} - \frac{2LU(\sin^2\vartheta - \cos^2\vartheta)(-3+\cos^2\vartheta)}{\cos^5\vartheta}\frac{1}{R^3} + \cdots\right]R^2 d\vartheta^2$$
$$- \left[1 + \frac{2L}{\cos^3\vartheta}\frac{1}{R^2} + \frac{2LU(-3+\cos^2\vartheta)}{\cos^5\vartheta}\frac{1}{R^3} + \cdots\right]\mathcal{C}^2 R^2\sin^2\vartheta\,d\phi^2 . \tag{B19}$$

Transforming into Bondi's coordinates $\{u, r, \theta, \phi\}$ by expansions

$$U = \overset{o}{\pi}(u,\theta) + \overset{1}{\pi}(u,\theta)/r + \cdots ,$$
$$R = q(u,\theta)\,r + \overset{o}{\sigma}(u,\theta) + \overset{1}{\sigma}(u,\theta)/r + \cdots , \tag{B20}$$
$$\vartheta = \overset{o}{\tau}(u,\theta) + \overset{1}{\tau}(u,\theta)/r + \overset{2}{\tau}(u,\theta)/r^2 + \cdots ,$$

we compare the metric expansion with Bondi's metric and restrict thus functions entering transformation (B20). We arrive at

$$\overset{o}{\pi} = \frac{u}{q} + \text{const}, \quad \overset{1}{\pi} = -\frac{q_{,\theta}^2}{q^3}\frac{u^2}{2} - \frac{2L\mathcal{C}}{q_{,\theta}\sin\theta - q\cos\theta},$$

$$q = \frac{\sin\theta}{2\mathcal{C}}\left[\left(\chi\frac{\sin\theta}{\cos\theta + 1}\right)^{\mathcal{C}} + \left(\chi\frac{\sin\theta}{\cos\theta + 1}\right)^{-\mathcal{C}}\right], \quad \chi = \text{const}, \qquad (B21)$$

$$\overset{o}{\sigma} = \frac{u}{2q}(q_{,\theta}^2 + q^2 - 1), \quad \overset{1}{\sigma} = \tfrac{1}{2}u^2\left(\frac{q_{,\theta}^2}{q^3} + qc_{,u}^2\right) + \frac{L\mathcal{C}(2q^2\mathcal{C}^2 - \sin^2\theta)}{(q^2\mathcal{C}^2 - \sin^2\theta)(q_{,\theta}\sin\theta - q\cos\theta)},$$

$$\overset{o}{\tau}_{,u} = 0, \quad \overset{o}{\tau}_{,\theta} = \pm\frac{1}{q}, \quad \overset{1}{\tau} = \pm\frac{q_{,\theta}u}{q^2}, \quad \overset{2}{\tau} = -\frac{q_{,\theta}}{q^4}(q_{,\theta}^2 + q^2 - 1)\frac{u^2}{2} + \frac{L\mathcal{C}\sin\theta}{q(q^2\mathcal{C}^2 - \sin^2\theta)},$$

and

$$a_1 = 0, \quad c = u\frac{\mathcal{C}^2 - 1}{2\sin^2\theta}. \qquad (B22)$$

We choose the + sign and the simplest unboosted Bondi's system with $\chi = 1$. Using the results above to transform the spacelike translational Killing vector $\partial/\partial z$, which in coordinates $\{t, \rho, z, \phi\}$ just reads

$$\eta^\alpha = [\,0,\ 0,\ a_0,\ 0\,], \qquad (B23)$$

into Bondi's coordinates, we find

$$\eta_u = a_0\left\{\frac{1}{2\mathcal{C}\sin^2\theta}[(\mathcal{C}^2 + 1)q_{,\theta}\sin\theta + (\mathcal{C}^2 - 1)q\cos\theta] + \frac{uc_{,u}^2}{\mathcal{C}}(q_{,\theta}\sin\theta - q\cos\theta)\frac{1}{r} + \cdots\right\},$$

$$\eta_r = a_0\left[\frac{1}{\mathcal{C}}(q_{,\theta}\sin\theta - q\cos\theta) - \frac{u^2 c_{,u}^2}{2\mathcal{C}}(q_{,\theta}\sin\theta - q\cos\theta)\frac{1}{r^2} + \cdots\right]$$

$$\eta_\theta = a_0\left\{\frac{\sin\theta}{\mathcal{C}q}(q_{,\theta}^2 - qq_{,\theta}\cot\theta + 1)r + \frac{uc_{,u}}{\mathcal{C}\sin\theta}[q_{,\theta}\sin\theta\cos\theta + q(\mathcal{C}^2 - \cos^2\theta)]\right.$$

$$\left. + \left[-\frac{u^2 c_{,u}^2}{2\mathcal{C}\sin\theta}(3\cos\theta(q_{,\theta}\sin\theta - q\cos\theta) - q\mathcal{C}^2) - \frac{2L\mathcal{C}^2\sin\theta}{(q^2\mathcal{C}^2 - \sin^2\theta)(q_{,\theta}\sin\theta - q\cos\theta)}\right]\frac{1}{r} + \cdots\right\}. \qquad (B24)$$

Comparing this with the asymptotic expansion of the general translational Killing vector (57), in which we put $b_0 = 0$, we obtain the agreement if the constant parameters

$$a_1 = 0, \quad f_0 = \frac{2La_0^4}{\mathcal{C}}, \quad g_0 = 0. \qquad (B25)$$

The above results generalize those obtained in [7] for the "averaged" time-symmetric cylindrical waves without a string along the symmetry axis. As emphasized in [7] in the case of these waves null infinity is smooth in all generic directions except those orthogonal to the axis. If the string is present, null infinity is not smooth also in the directions of the string, $\theta = 0, \pi$.